\documentclass[journal]{IEEEtran}
\usepackage{xcolor}
\usepackage{cite}
\usepackage{amsmath}
\usepackage{multirow}
\usepackage{graphicx}
\usepackage[switch]{lineno}

\begin{document}

\title{Effect of skin-to-skin contact on stimulation threshold and dosimetry}

\author{Essam A. Rashed,~\IEEEmembership{Senior Member,~IEEE,} Yinliang Diao,~\IEEEmembership{Member,~IEEE,} Shota Tanaka, Takashi Sakai, Jose Gomez-Tames,~\IEEEmembership{Member,~IEEE,} Akimasa Hirata,~\IEEEmembership{Fellow,~IEEE}

\thanks{This study is organized under the activity of task force of IEEE International Committee on Electromagnetic Safety (Task Force~2 under Subcommittee~6). The authors would like to thank Dr. W. P. Segars (Duke University) for providing us with the XCAT phantom and Dr. J. P. Reilly (Metatec Associates) for useful discussion on the sensation threshold in measurement. E. A. Rashed is with the Department of Mathematics, Faculty of Science, Suez Canal University, Ismailia 41522, Egypt. Y. Diao is with the South China Agricultural University, China. S. Tanaka, T. Sakai, J. Gomez-Tames and A. Hirata are with the Department of Electrical and Mechanical Engineering, Nagoya Institute of Technology, Nagoya, Japan. Corresponding author e-mail: erashed@science.suez.edu.eg}}

% The paper headers
%\markboth{IEEE TRANSACTIONS ON ELECTROMAGNETIC COMPATIBILITY}%
{ }

\maketitle

\begin{abstract}
The human dosimetry for electromagnetic field exposure is an essential task to develop exposure guidelines/standards for human safety as well as product safety assessment. At frequencies from a few hundred Hz to 10 MHz, the adverse effect to be protected is the stimulation of the peripheral nervous system. The \emph{in~situ} electric field in the skin is used as a surrogate of nerve activation. In the low-frequency dosimetry, a high but inaccurate \emph{in~situ} electric field has been reported at positions where a skin-to-skin contact exists, whose relation to the stimulation is controversial. One of the reasons for high electric fields may be attributable to the current resolution of anatomical models. 

In this study, we first evaluate the stimulation threshold at postures of skin-to-skin contact experimentally for different hand/finger positions to represent skin touching/non-touching scenarios. We confirm that the skin-to-skin contact does not lower the threshold current of magnetic stimulation devices needed to induce pain. Second, a new method is proposed for hand modeling to configure different finger positions using static hand models with similar postures to the experiments. We compute the \emph{in~situ} electric field at skin-to-skin contact for the different hand posture scenarios that indicate an excessive raise of the electric field in skin-to-skin regions that is not justified by the experiments. The comparison suggests that a high \emph{in~situ} electric field in the skin would be caused by poor modeling of the skin layers, which is not enough to represent in a resolution of the order of a millimeter. This skin-to-skin contact should not be considered to set the restriction in the international exposure guidelines/standards as well as product safety assessment.

\end{abstract}

\IEEEpeerreviewmaketitle

%--------------------------------------
% I. Introduction
%--------------------------------------

%\linenumbers
\section{Introduction}

International exposure guidelines/standards have been developed to provide human protection from electromagnetic fields. There exist two international exposure guidelines/standards, which are mentioned in World Health Organization (WHO) documents; International Commission on Non-Ionizing Radiation Protection \cite{ICNIRP1998, ICNIRP2010} and IEEE International Committee on Electromagnetic Safety (ICES) Technical Committee 95 \cite{IEEEC9512019}. These guidelines/standards are classified into two regimes based on the adverse health effects; electrical stimulation and heating below and above approximately 100 kHz, respectively, for continuous exposures. The physical quantities prescribed in the guidelines/standards are thus variant for different effects; for local/non-uniform exposures, \emph{in situ} electric field averaged over specific volume/line for stimulation and specific absorption rate (SAR) averaged over 10 g of tissue for heating.

In the dosimetry for low-frequency (LF) exposure, the \emph{in~situ} electric field averaged over small volume/short length is used; 2~$mm$ cube in the ICNIRP guidelines and 5~$mm$ line average in the IEEE C95.1 standard. To compute such internal field strength for field exposure, an electromagnetic model imitating the human is needed. In the IEEE C95.1 standard  \cite{IEEEC9512019}, the ellipsoid is used to relate the external magnetic field and \emph{in~situ} electric field strength. In the ICNIRP guidelines \cite{ICNIRP2010}, anatomically-realistic human body models were considered to relate the internal and external field strength. Then, a procedure for evaluating the \emph{in~situ} electric field is mentioned as follows

``\emph{As a practical compromise, satisfying requirements for a sound biological basis and computational constraints, ICNIRP recommends determining the induced electric field as a vector average of the electric field in a small contiguous tissue volume of $2\times2\times2~mm^3$. For a specific tissue, the $99^{th}$ percentile value of the electric field is the relevant value to be compared with the basic restriction}". 

The above-mentioned difference suggests that unlike the ellipsoid, there exist several issues to be resolved in the computation with anatomical models. The $99th$~\%ile value of \emph{in~situ} electric field/current density and/or their spatial averaging are introduced to consider the numerical artefacts. The metric of the $99th$ \%ile value is first introduced for the uniform electric field exposures \cite{Hirata2001TBE}. Thus, it is not directly applied to the non-uniform exposure. These combined topics are listed as issues to be resolved in the IEEE ICES research agenda \cite{ICES_Research_Agenda}.

There are several causes related to the numerical artefacts. A human model is segmented into more than dozen tissues (over 50 tissues in the whole body), and then an identical electrical conductivity is assigned to each tissue; the same value is assigned even to different body parts. In addition, the human body model itself is discretized in small voxels (finite-difference method) or tetrahedrons (finite element method). A resolution of the discretization is typically the order of a millimeter. Then, in the finite-difference methods, which is more commonly used, the stair-casing error cannot be ignored \cite{Laakso2012PMB, GomezTames2017TEMC}. In particular, the skin is composed of multiple layers, each of which is insufficient to represent in a resolution of a few millimeters \cite{Rashed2019Access, Schmid2016PMB}. In addition, due to the finite resolution, the skin-to-skin contact in the model is also suggested \cite{DeSantis2015BPEE, Li2015Bio}. The equivalent conductivity of tissues is discussed in \cite{Kavet2015HP, DeSantis2015BPEE}. Thus, even if we use any computational methods, it is difficult to estimate the \emph{in situ} electric field in details without any error.

De Santis et al., 2016 \cite{DeSantis2016PMB} suggested exposure scenarios and posture, in which the \emph{in~situ} electric field is enhanced. However, this study just presented the \emph{in~situ} electric field for specific cases. The international guidelines/standards do not consider extreme cases in ICNIRP radio frequency (RF) guidelines and also mentioned as ``\emph{The safety limits for electrostimulation are based on conservative assumptions of exposure. However, they cannot address every conceivable assumption}" in IEEE C95.1-2019. A question arises if potential enhancement of computed electric field, which may be attributable to a limitation of skin modeling, is related to the electrostimulation.

%-----------------------
% Fig-1
\begin{figure}
\centering
\includegraphics[width=0.5\textwidth]{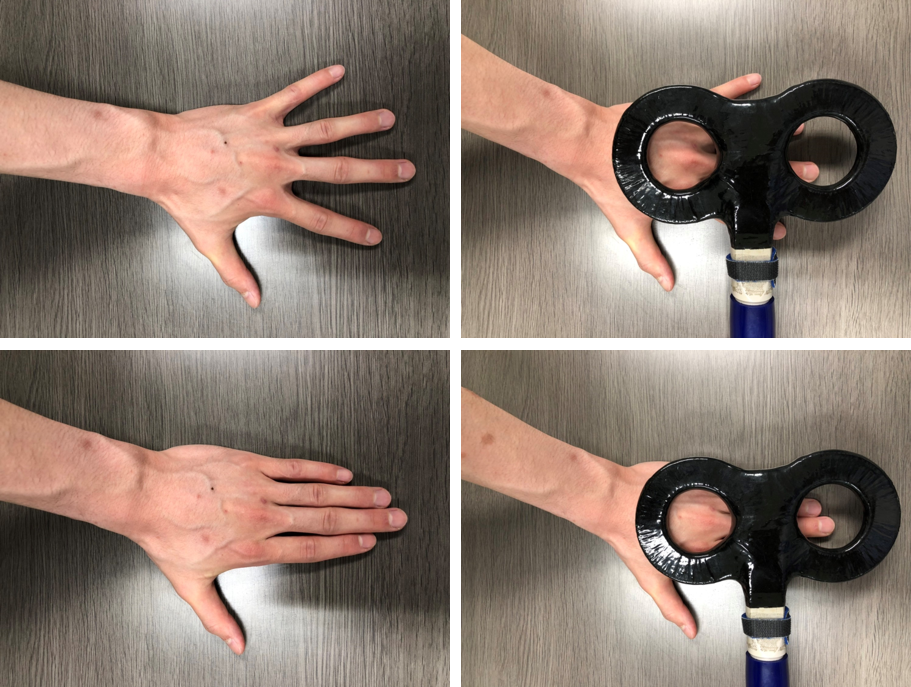}
\caption{Magnetic stimulation experiment setup for the first scenario with open hand mode (up) and close hand mode (bottom). Hand posture (left) and coil position (right) are shown for clarifications.}
\label{fig01}
\end{figure}
%-----------------------

At least three studies presented experimental results which could be potentially related to the threshold change of the electrostimulation by skin-to-skin contact. However, the description of the observation is sketchy, and there are several possible confounders to derive conclusions therein. The confounders are summarized as three items: i) the contacting-hand condition would involve a large induction loop that would be absent when the hands are separated, ii) the hand contact is described as ``touching" in \cite{DeSantis2015BPEE} and ``clasping" in \cite{Schmid2016PMB}, iii) with contacting hands, \cite{DenBoer2002JMRI} did not report exactly where the sensation was felt; fingers or wrists \cite{Reilly2018private}.

Therefore, systematic experimental and computational evaluation are needed to assess the change of the stimulation threshold, which is attributable to the skin-to-skin contact. In this study, the experiment to assess potential change of stimulation threshold has been conducted at the skin-to-skin contact in fingers. We then develop new hand/finger models to investigate the skin-to-skin contact effect computationally for computational replication.

%-----------------------
% Fig-2
\begin{figure}
\centering
\includegraphics[width=0.5\textwidth]{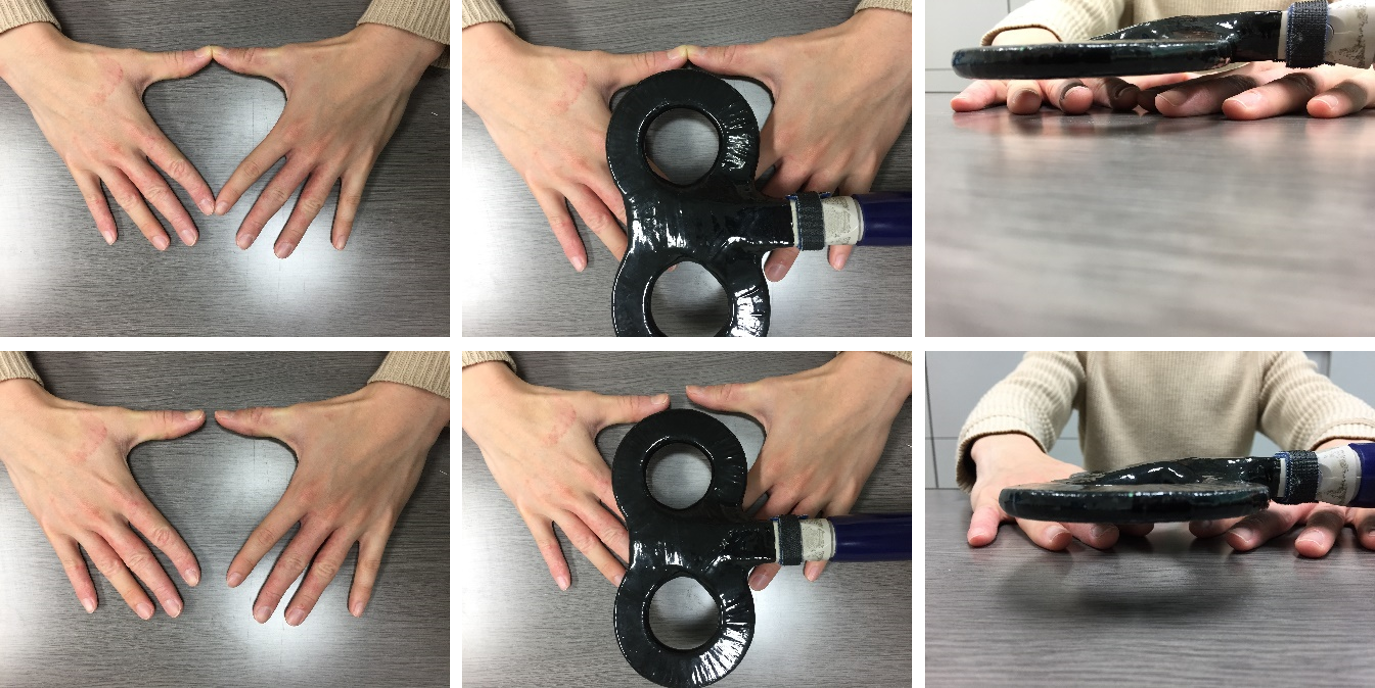}
\caption{Magnetic stimulation experiment setup for the second scenario with close loop mode (up) and open loop mode (bottom). Hand postures (left), coil position from up (middle) and frontal (right) views are shown for clarifications.}
\label{fig02}
\end{figure}
%-----------------------

%-----------------------
% Fig-3
\begin{figure*}
\centering
\includegraphics[width=\textwidth]{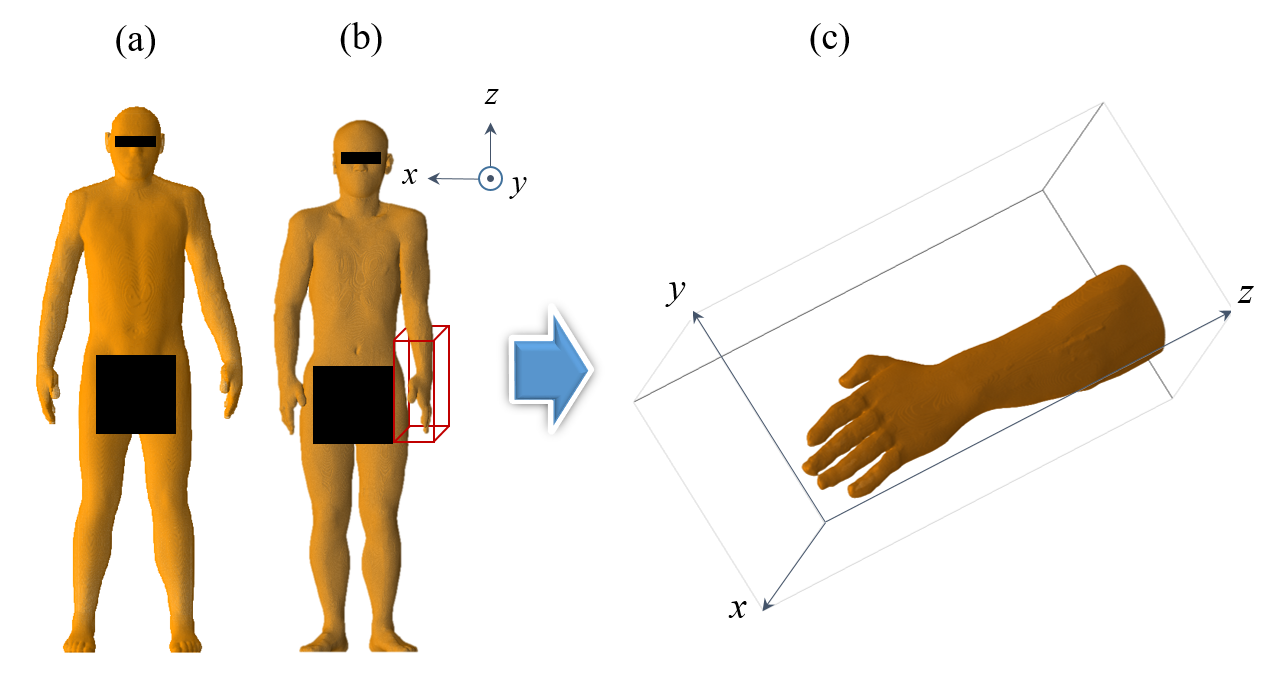}
\caption{(a) XCAT and (b) TARO models in frontal standing position. Left forearm of TARO model labeled with red box is extracted in (c).}
\label{fig03}
\end{figure*}
%-----------------------

%-----------------------
% II. Measurements protocol
%-----------------------

\section{Measurements protocol}

Stimulus threshold was measured for different skin-to-skin conditions of hands using magnetic stimulation. The magnetic stimulator\footnote{Rapid square, Magstim Co., England}, with a figure-eight designed coil (outer diameter of 70~$mm$), was used to deliver a pulse to the hand of the subject. A group of eight male volunteer subjects were participating in the study (22.5 $\pm$1.07 years old). All subjects were confirmed to satisfy the criteria for experiment participation. That is, they did not have a pacemaker, any breaks in the skin under the area where the coil planned to be placed on, and no known neurologic or musculoskeletal pathologies affecting the hand to be tested. All subjects were awake, sat comfortably in a reclining chair within a quiet environment, and were requested to place the left hand on a table and remain relaxed with their eyes closed. Prior to the experiment, all subjects are sure that hands are completely dry with no sweating. The magnetic stimulation was performed considering two scenarios: 1) center of the figure-eight coil is located over the middle finger of the left hand, and 2) center of the coil is located over the tip of index fingers of both hands. The hand(s) posture and coil orientation are shown in Figs.~\ref{fig01} and \ref{fig02} for the first and second scenarios, respectively. In the first scenario, two modes are considered: fingers abduction (open hand mode) and fingers adduction (close hand model). In the second scenario, two modes are also considered. Close loop mode, where thumb and index fingers of both hands are tip-touched to form a shape of closed loop. Open loop mode, where thumb and index fingers are close to each other without being touched (distance is approximately 10 $mm$).

All subjects were asked to verbally answer whether they felt a sensation, and the maximum stimulation output (MSO) of the device was then recorded without notifications to the subject. Although the minimum sensation is inevitably subjective between volunteer subjects, the threshold comparison between conditions (contact versus no contact) for the same subject is not (or at least less subjective). During the experiment, subjects were reminded that they could withdraw if they wished at any stage and that the level of discomfort should be one that they believed repeatable and not the absolute maximum they could tolerate. The stimulus threshold was defined as the lowest stimulation intensity capable of generating the minimal sensation following the method of limits, with three ascending trials and three descending trials. Each ascending trial was followed by a descending trial for a total of six trials per subject. For ascending trials, MSO of the coil device was increased from 0\% in increments of 1\% until stimulus was perceived and verbally reported by the subject. For descending trials, the MSO was increased 10\% beyond the previous ascending threshold, and MSO was decreased by 1\% until stimulus was perceived. The mean over the six trials was taken as the stimulus threshold. This protocol was applied in the first session for stimulation threshold detection for open hand mode. In a second session, the protocol was repeated for close hand mode. The rest time between trials and sessions was of 30 and 120 seconds, respectively. The same scenario was repeated for open and close loop modes.

%-----------------------
% III. Computational hand models
%-----------------------

\section{Computational hand models}

Anatomical models representing human hand is used to conduct the dosimetry for the figure-eight coil. Currently available models are commonly provided in a single static hand posture that requires customization to fit with the above-mentioned scenarios. Two anatomical models are used in this study; the Japanese adult male model (TARO) developed at NICT, Japan \cite{Nagaoka2004PMB} and 4D extended cardiac-torso (XCAT) phantom developed at Duke University, USA \cite{Segars2010MP}. The anatomy of human includes 27 bones in the hand, where 19 forming fingers and palm. The bone structures are connected by joints with different rotation space and degree of freedom, which make it complicated modeling problem \cite{Erol2007CVIU}. In the context of computer graphics and animation, several approaches are presented for hand modeling and simulation \cite{Wheatland2015CGF}. However, these approaches did not pay enough attention to model skin-to-skin contact in a reliable manner. Therefore, we present a simple, though effective, approach that can be used to adjust finger position to simulate realistic situations where skin-to-skin contact is more emphasized. To adjust hand/finger positions, we use kinematic hand modeling with fixed and free-move joint points as detailed in Section III.C below. This model fits the hand posture required for this study as it allows minimal anatomy changes between different skin-to-skin contact positions.

%-----------------------
% Fig-4
\begin{figure*}
\centering
\includegraphics[width=\textwidth]{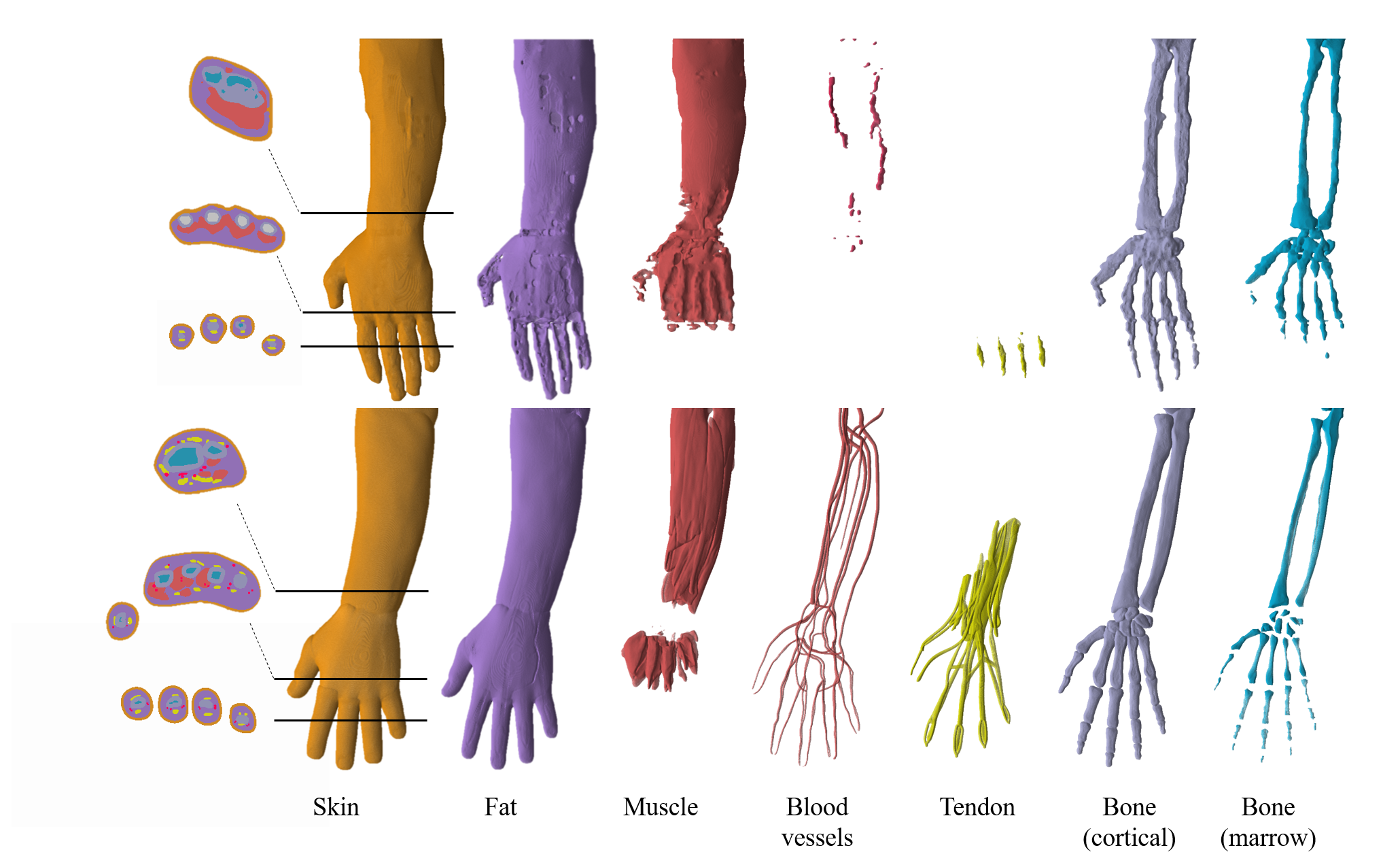}
\caption{Volume rendering of different anatomical tissues (left forearm) in TARO (top) and XCAT (bottom) models. Far left is a selected cross-sectional slices of both models. This demonstrate that XCAT model provide a more detailed and realistic anatomical structures.}
\label{fig04}
\end{figure*}
%-----------------------

%-----------------------
% A. TARO hand model
%-----------------------

\subsection{TARO hand model}

TARO model \cite{Nagaoka2004PMB} was frequently used in evaluation of electromagnetic dosimetry.  Original TARO model is of $2.0~mm$ resolution, which is relatively low for high accuracy computation considering the effect of the skin. A higher resolution model of $0.5~mm$ is generated using the method detailed in \cite{Rashed2019Access, Taguchi2018Toyama}. A forearm region of both right and left hands is extracted as shown in Fig.~\ref{fig03}. It is expected that electric field will be highly induced in the region of fingers. However, it is difficult to exactly estimate the minimum required volume for this evaluation. This is because the induced field is characterized by Faraday’s law and thus non-finger region may also contribute to the induced electric field in the finger. Therefore, a complete forearm is used in this study to avoid any potential truncation effect. The TARO hand consists of seven tissues, skin, muscle, fat, blood vessels, tendon, bone (cortical), and bone (marrow) as demonstrated in Fig.~\ref{fig04}.

%-----------------------
% B. XCAT hand model
%-----------------------

\subsection{XCAT hand model}

XCAT model \cite{Segars2010MP} is designed for high-quality medical imaging simulations such as those used to simulate CT, PET and SPECT imaging. It is hardly being used in electromagnetic dosimetry studies. As the XCAT phantom is provided as a software, it is possible to be customized. We have generated a $0.5~mm$ model with skin thickness of $2.0~mm$ (default parameters ignore skin tissues). The X-ray attenuation model is then transferred into labeled model that represents different anatomical tissues. Forearms of XCAT phantom are extracted similar to TARO model (Fig.~\ref{fig03}). To demonstrate the difference in anatomy structure representation in the two models, a volume rendering of different tissues in left forearm is shown in Fig.~\ref{fig04}.

%-----------------------
% Fig-5
\begin{figure*}
\centering
\includegraphics[width=\textwidth]{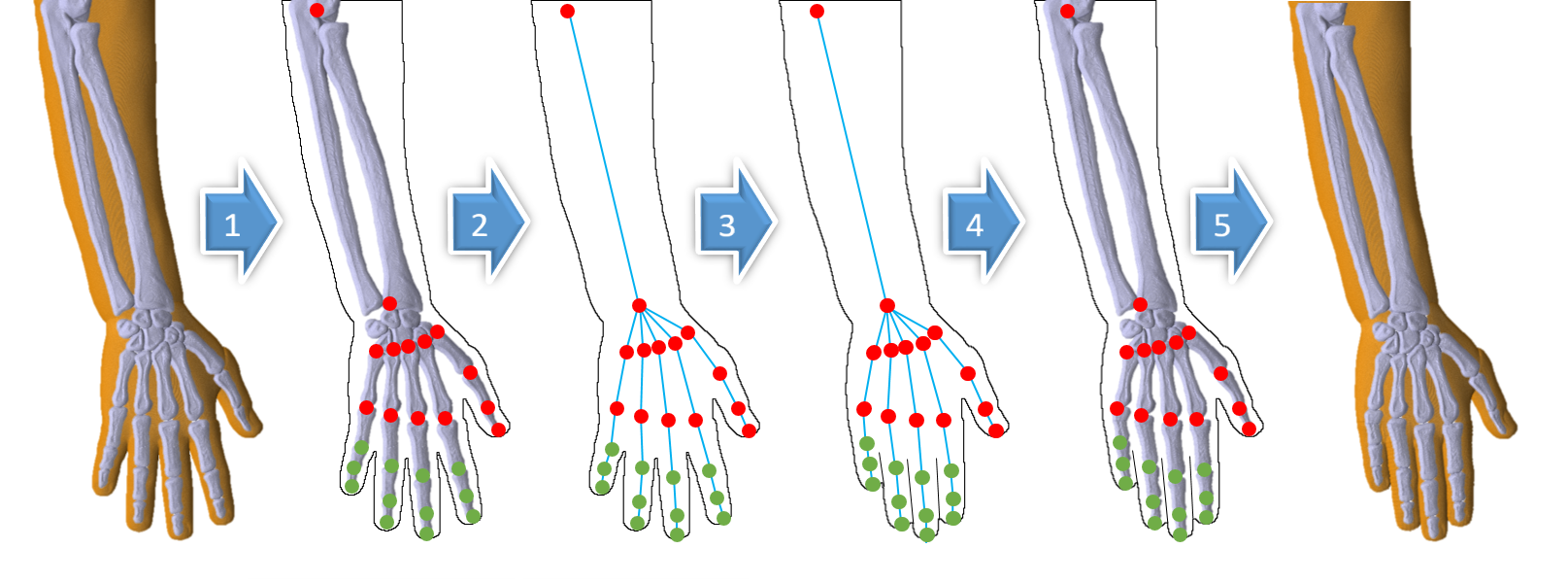}
\caption{Illustration of the forearm modeling steps from open to close hand modes. (1) Cortical bones are extracted then hand kinematic joint locations and finger tips are labeled as vertices. Green and red color labels indicate free-move and fixed vertices for finger positing, respectively. (2) Edge connections (blue lines) are presented to connect vertices and generate forearm skeleton. (3) Free-move vertices are shifted to new positions for different hand posture in 3D space. Details of vertices positioning are in Fig.~\ref{fig06}. (4) Cortical bones are registered to the new forearm skeleton. (5)~Other hand tissues are registered to the new bone posture.}
\label{fig05}
\end{figure*}
%-----------------------

%-----------------------
% Fig-6
\begin{figure}
\centering
\includegraphics[width=.5\textwidth]{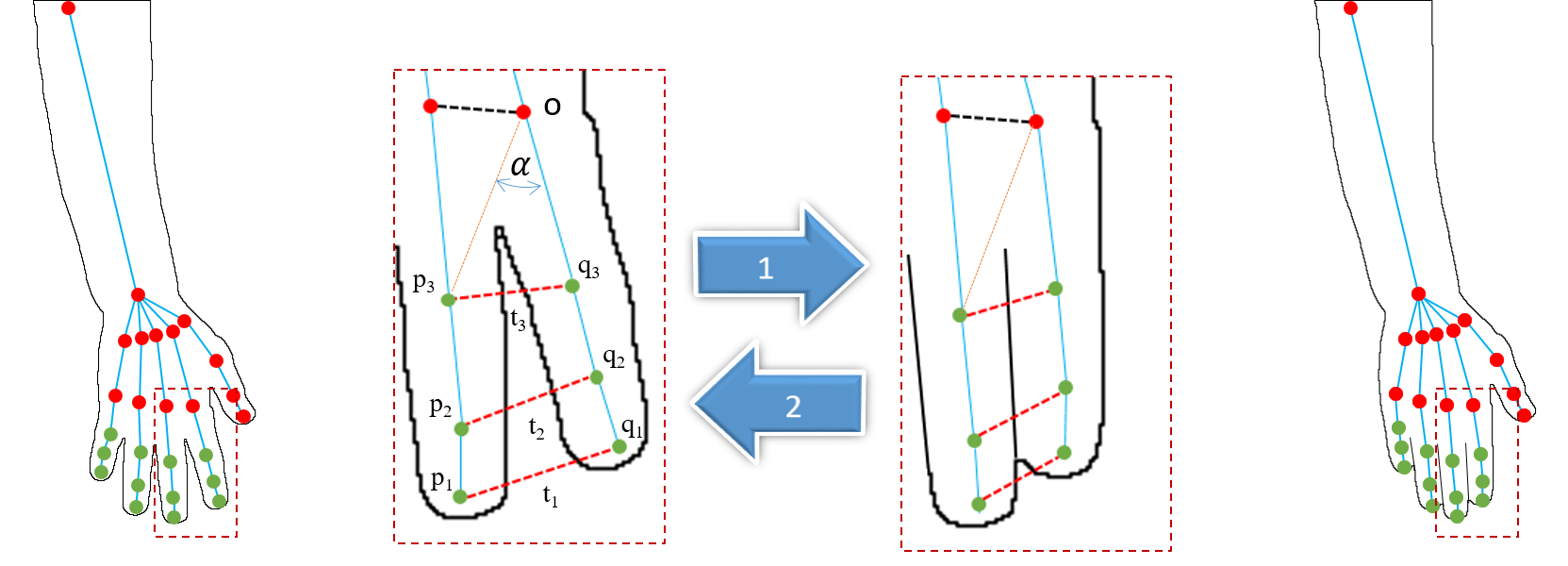}
\caption{Finger positioning from open to close (Direction~\#1) hand modes and vice versa (Direction~\#2). Magnified regions corresponding to middle and index fingers are shown in the middle.}
\label{fig06}
\end{figure}
%-----------------------

%-----------------------
% C. Model customization (first scenario)
%-----------------------

\subsection{Model customization (first scenario)}

To simulate the hand postures corresponding to the first scenario, the hand models are customized to generate different finger positions. The finger positioning process is conducted in five steps (Fig.~\ref{fig05}) to simulate a realistic human hand movement. 

\begin{itemize}
\item {\bf Step 1:} The kinematic joint and fingertips locations of the hand cortical bones are labeled as vertices. Vertices connected to radius, carpus, and metacarpus bones are assumed to be stationary. Other vertices are labeled for free-move.
\item {\bf Step 2:} Vertices are connected with edges representing bone structures to generate hand skeleton.
\item {\bf Step 3:} 
\begin{itemize}

\item In case of close hand mode, the new position of $q_i$ finger is found by solving the following equation
\begin{equation}
\tilde{q}_i=\arg \min_{q_i} t_i~\textnormal{subject to} ~p_{s} \cap q_{s}=\phi ~~~ \forall i,
\end{equation}
\begin{equation*}
t_i=dist(p_i,q_i) 
\end{equation*}
where $dist(.,.)$ is the Euclidean distance, $p_{s}$ and $q_{s}$ represent the skin volume of the fingers $p$ and $q$, respectively. In other words, the $q_i$ vertices are moved towards the corresponding $p_i$ ones such that no skin overlap occurs. Illustration indicate vertices positing approach is shown in Fig.~\ref{fig06} (Direction~\#1). This process is repeated for all figures accordingly.

\item In case of open hand mode, one finger is assumed to stand stationary ($p_i$, $i=1,...,3$) and the second finger ($q_i$, $i=1,...,3$) is moved away. The new position ($\tilde{q}_i$) is computed as:

\begin{equation}
\tilde{q}_i=\left[ \begin{array}{ccc} 1 & 0 & 0 \\ 
0 & \cos(\alpha)  & -\sin(\alpha)\\
0 & \sin(\alpha)& \cos(\alpha) \end{array} \right] q_i,
\label{eq02}
\end{equation}

where $q_i$ is presented in the Cartesian grid with origin point $o$ as shown in Fig.~\ref{fig06} (Direction~\#2), $\alpha=\alpha_\circ+\Delta$ is the angle of the new position, $\alpha_\circ$ is the original angle and $\Delta$ is the rotation angle. The rotation equation in Eq.~(\ref{eq02}) considers the movement counterclockwise on y-z plane only (for simplicity). Extension to 3D rotation is simple and direct. This process is repeated for all fingers accordingly.
\end{itemize}

\item {\bf Step 4:} Cortical bones corresponding to the new hand posture are generated using 3D control-point-based non-rigid registration \cite{Allen2003TOG}.

\item {\bf Step 5:} Other hand tissues are registered to cortical bones using non-rigid registration.
\end{itemize}

%-----------------------
% D. Model customization (second scenario)
%-----------------------

\subsection{Model customization (second scenario)}

\begin{figure}
\centering
\includegraphics[width=.5\textwidth]{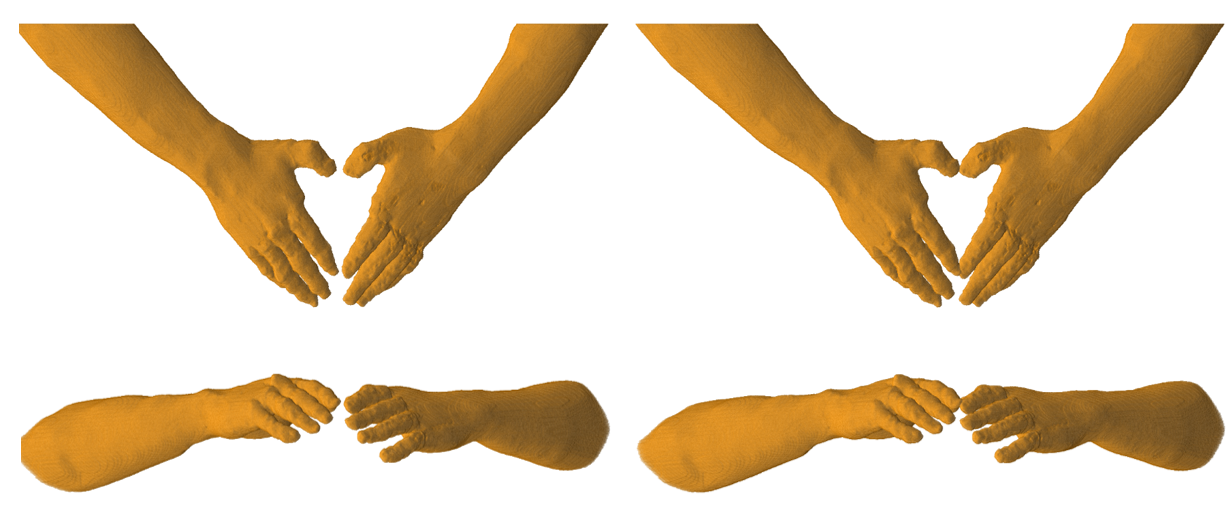}
\caption{TARO hand positions for the second scenario forming open (left) and close (right) loops. Frontal position is shown below the top position.}
\label{fig07}
\end{figure}

\begin{figure}
\centering
\includegraphics[width=.4\textwidth]{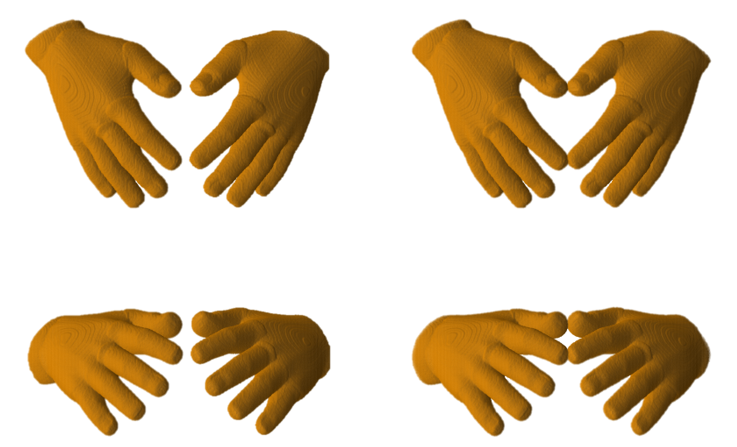}
\caption{XCAT hand positions for the second scenario forming open (left) and close (right) loops. Frontal position is shown below the top position.}
\label{fig08}
\end{figure}

The second scenario considers an open/close loop formed by the thumb and index fingers of both hands. The original forearms extracted from the original models (both TARO and XCAT) are rotated in 3D space such that the tips of thumb and index fingers are in touching position as shown in Figs.~\ref{fig07} and~\ref{fig08}. The original anatomical structure and finger's positioning does not change. The non-touching position is then generated by shifting the left hand approximately 10~$mm$ away from the right hand.

%-----------------------
% E. Electromagnetic modeling
%-----------------------

\subsection{Electromagnetic modeling}

The magnetic vector potential is computed as a source for the figure-eight coil with a single loop (70~$mm$ in diameter) for each winding. The electric field was calculated for a sinusoidal wave with fixed frequency at 10~kHz using 1~A of injection current in the coil. The commercial software COMSOL \footnote{COMSOL Multiphysics \textsuperscript{ \textregistered} v. 5.4, COMSOL AB, Stockholm, Sweden} was used to obtain the magnetic vector potential by finite element method. The induced scalar potential is obtained in a voxelized volume conductor model of the hands by the following equation:

%Eq-3
\begin{equation}
 \nabla \left[ \sigma \left( -\nabla \phi - j \omega A_0 \right) \right]=0,
\label{eq_isp}
\end{equation}

where $A_0$ and $\sigma$ denote the magnetic vector potential of the applied magnetic field and tissue conductivity, respectively. This equation assumes the magneto-quasi-static approximation (applicable in the 10 kHz frequency band of magneto-stimulator), i.e., the displacement current is negligible when compared to the conduction current, and the external magnetic field is not perturbed by the induced current in the hand tissues \cite{Hirata2013PMB}. The isotropic conductivity values of different tissues are listed in Table~\ref{Tab1}. The scalar potential finite difference method \cite{Dawson1996AC} was used to numerically solve the Eq. (3), which was discretized in a sparse matrix equation. The matrix equation was solved iteratively using the geometric multigrid method with successive over-relaxation \cite{Laakso2012PMB}. The number of multigrid levels was six, and the iteration continued until the relative residual was smaller than $10^{-6}$.

%-----------------------
% Table 1
%-----------------------

\begin{table}
\centering
\footnotesize
\caption{Human tissue conductivity ($\sigma_n$) of TARO/XCAT models.}
\label{Tab1}
\setlength{\tabcolsep}{3pt}
\begin{tabular}{| l|l | c| l|l | c|}
\hline
$n$ &{\bf Tissues} &{\bf $\sigma_n$}& $n$ &{\bf Tissues} &{\bf $\sigma_n$}\\
\hline
\hline
1 &Skin & 0.1 & 5 &Bone (cortical) & 0.02 \\
2 &Fat &  0.04 & 6 &Tendon &0.28 \\
3 & Muscle &  0.35 & 7 &Blood & 0.7 \\
4 & Bone (marrow) & 0.05 &&&\\
\hline 
\end{tabular}
\end{table}

%-----------------------
% IV. Results
%-----------------------

\section{Results}

%---------------------------------------------
% A. Measured Threshold for magnetic stimulations
%---------------------------------------------

\subsection{Measured threshold for magnetic stimulations}

\begin{figure*}
\centering
\includegraphics[width=\textwidth]{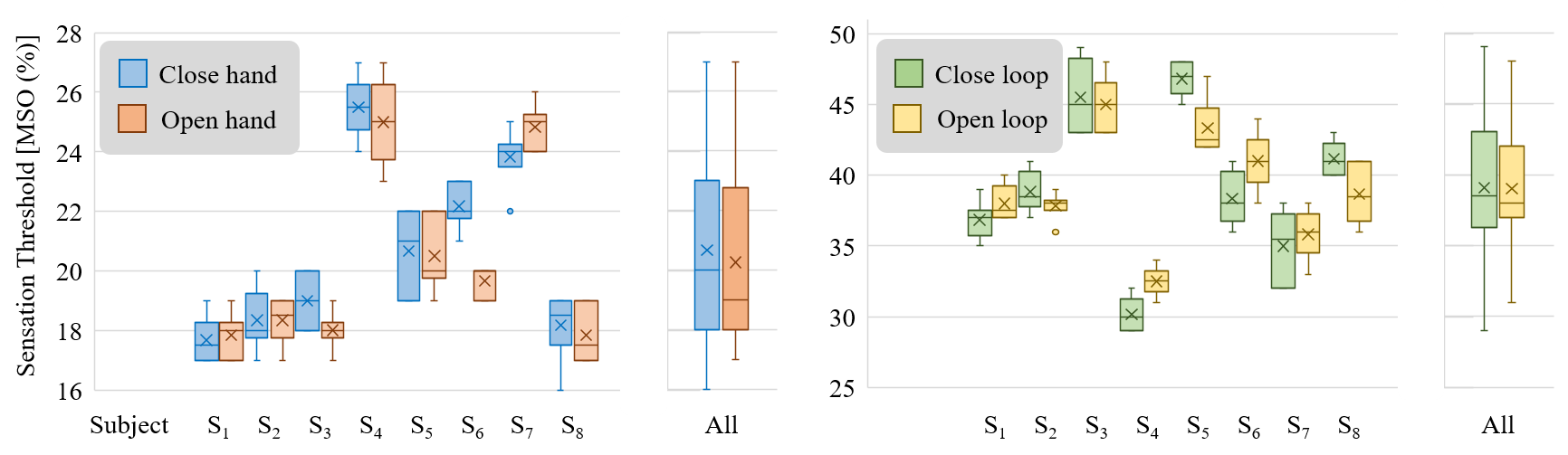}
\caption{Boxplot of sensation threshold measurements for the eight subjects ($S_1$-$S_8$). Left is results corresponding to first scenario (Fig.~\ref{fig01}) and right is results for second scenario (Fig.~\ref{fig02}). Cross indicates the mean value, center line indicates the median, box indicates first and third quartiles, whiskers are the maximum and minimum values.}
\label{fig09}
\end{figure*}

The measurements of threshold coil current (\% MSO) for different hand postures and position scenarios are shown in Fig.~\ref{fig09}. For the first scenario, a minimum difference of average threshold of 0.0\% in subject $S_2$ and a maximum difference of 11.3\% in subject $S_6$. However, the global average difference is 2.0\% with median MSO values of 19 and 20, for open and close hand postures, respectively. It is also clear from Fig.~\ref{fig09} that the maximum, minimum, first and third quartile values are of high consistency. In the second scenario, a minimum difference of average threshold of 1.1\% in subject $S_3$ and the maximum difference of 7.7\% in subject $S_5$. However, global average and median differences are 0.0\% and 0.01\%, respectively. A global threshold difference in the two scenarios was also observed. This comportment is expected because the stimulation region in the first scenario is higher than that in the second scenario, which may be attributable to more fibers ending. These measurements indicate a small difference in sensation threshold values in open and close hand/loop positions, suggesting marginal effects of the skin-to-skin contact on electrostimulation threshold.

%-----------------------
% Fig-7
%-----------------------

%-------------------------------------------
% B. Electric field distribution in hand models for figure-eight coil
%------------------------------------------

\subsection{Electric field distribution in hand models for figure-eight coil}

\begin{figure*}
\centering
\includegraphics[width=\textwidth]{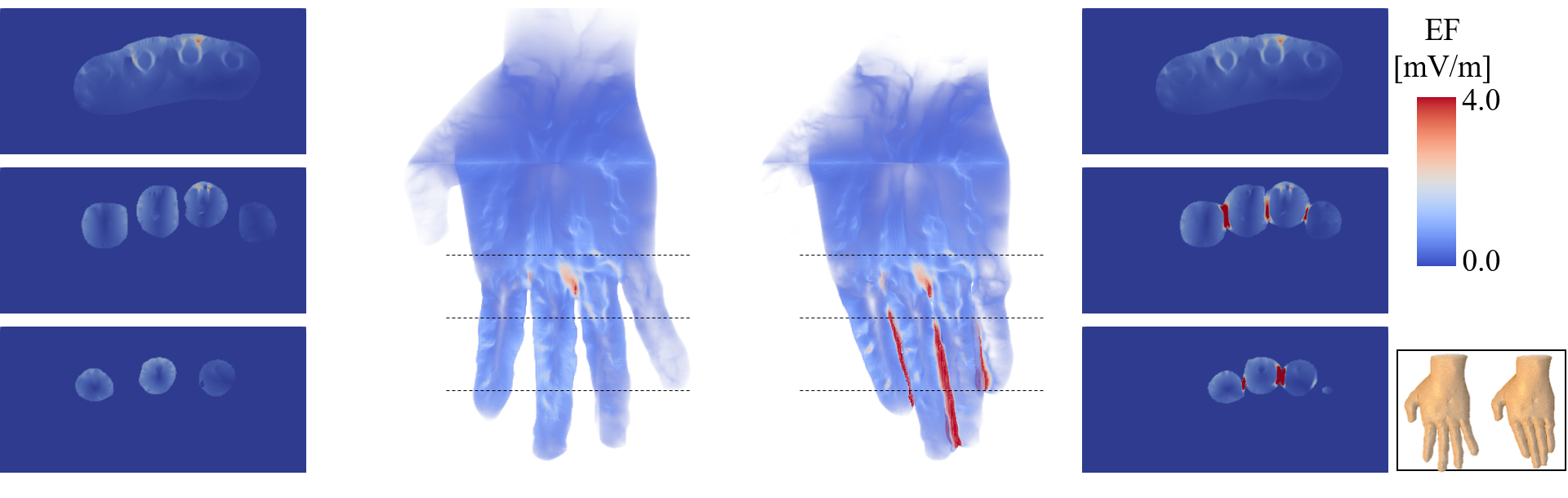}
\caption{Electric field distribution in TARO left hand in open (left) and close (right) positions for exposure to the magnetic field from the coil. A cross section slices of open and close hands in different positions (dashed lines) are shown on left and right sides, respectively. Hand anatomical positions are shown on bottom right side for reference. Significant electric field value is observed in skin-to-skin regions along finger contacts, whereas the electric field is almost identical in hand palm.}
\label{fig10}
\end{figure*}

\begin{figure*}
\centering
\includegraphics[width=\textwidth]{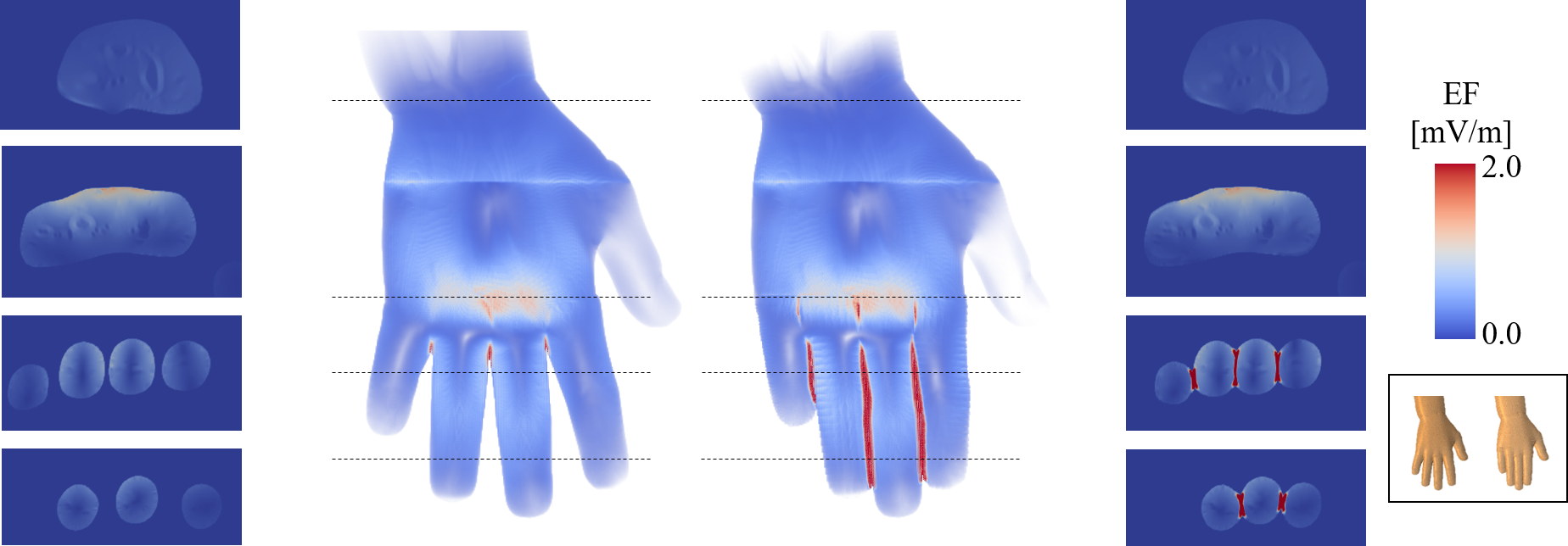}
\caption{TMS EF distribution in XCAT right hand in open (left) and close (right) positions. A cross section slices of open and close hands in different positions (dashed lines) are shown on left and right sides, respectively. Distribution of electric field is similar to the one in Fig.~\ref{fig10}.}
\label{fig11}
\end{figure*}

\begin{figure*}
\centering
\includegraphics[width=\textwidth]{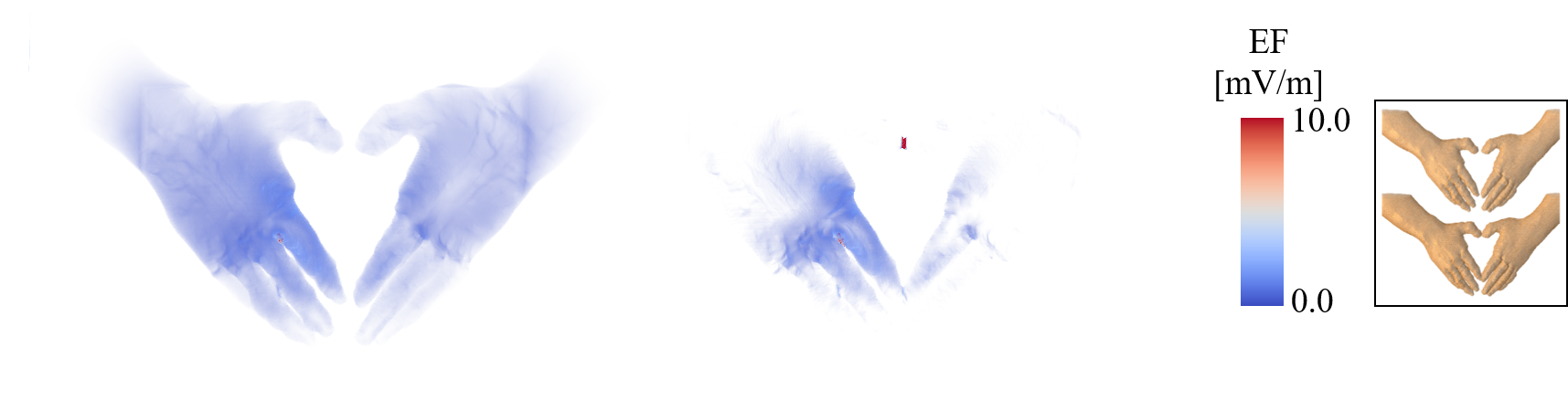}
\caption{Electric field distribution in the TARO of open (left) and close (right) loop positions.}
\label{fig12}
\end{figure*}

\begin{figure*}
\centering
\includegraphics[width=\textwidth]{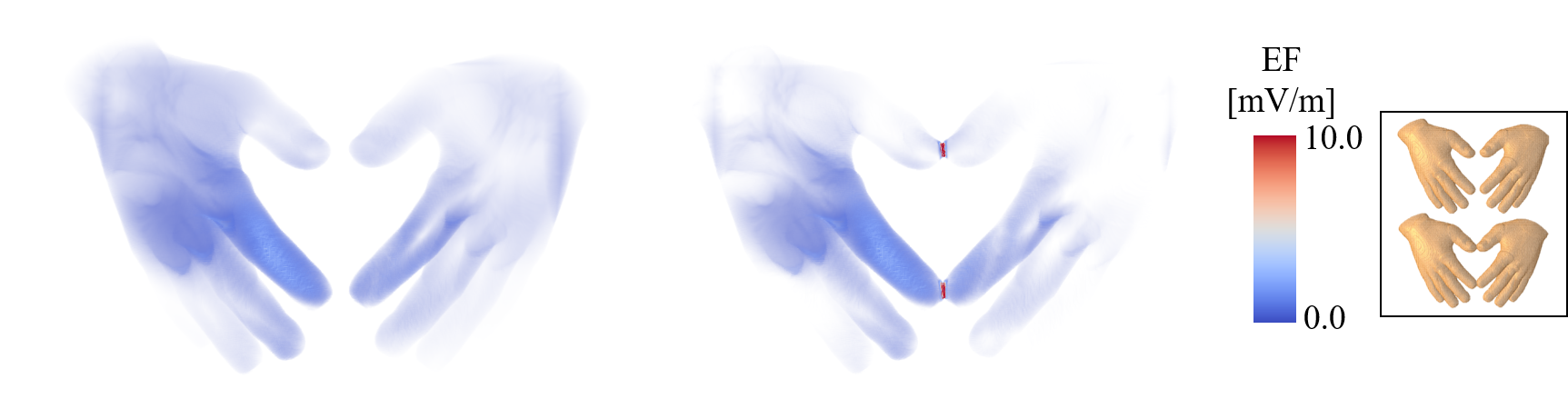}
\caption{Electric field distribution in the XCAT of open (left) and close (right) loop positions.}
\label{fig13}
\end{figure*}

For the first scenario, Figs.~\ref{fig10} and \ref{fig11} demonstrate a comparison of \emph{in situ} electric field in the hand model with/without the fingers contacted in TARO and XCAT hand models, respectively. For the second scenario, Figs.~\ref{fig12} and \ref{fig13} show the difference in the \emph{in situ} electric field in the hand model for two fingers formed open/close loop in TARO and XCAT hand models, respectively.

The spatial distributions of the electric field do not change for open and close conditions except in the skin-to-skin contact region. Also, the maximum values of the electric fields (excluding skin-to-skin contact regions) corresponded to the regions where sensation was felt by the subjects. Some variability of the electric field is confirmed due to subject-dependent anatomical factors and segmentation quality of the hands.

%------------------------------------
% B.1 First scenario
%------------------------------------

\subsubsection{First scenario}
Table~\ref{Tab2} shows the \emph{in situ} electric field at the 99.0, 99.9, and 100~\%ile values for the two hand models in different postures. The values are presented in terms of all tissues, skin only and non-skin tissues (others). From these results, a significant difference of the electric field is observed with different hand postures especially in the skin region for TARO model. It is relatively 7 (8) times larger at 99.0 (99.9)~\%ile considering both 2~$mm$ cube average and 5~$mm$ line average. The difference is relatively weaker in non-skin tissues with almost consistency at 99.0~\%ile (2 $mm$ average). In XCAT model, more stable values were observed in the skin tissue with a raise of approximately 4-5 times at different percentiles. For non-skin tissues, again, no (or small) difference is observed at 99.0~\%ile (2 $mm$ average) and 2-3 times higher at higher percentiles. From these data, we can also confirm a similar behavior in both TARO and XCAT models.

%-----------------------
% Fig-8
%-----------------------

%-----------------------
% Fig-9
%-----------------------

%-----------------------
% Fig 10
%-----------------------

%-----------------------
% Fig-11
% \\altair-ws\share\staff\Essam\Hand_Modeling\XCAT\0.5mm\simulation_results\
% E_open_320_500_600.raw
% E_close_320_500_600.raw
%-----------------------

%-----------------------
% Fig-12
%-----------------------

%-----------------------
% Fig-13
%-----------------------

%------------------------------------
% B.2 Second scenario
%------------------------------------

\subsubsection{Second scenario}
The computed electric field in the second scenario is shown in Table~\ref{Tab3} at different percentiles. For TARO model, the electric field is consistent in 99.9 and 99.0~\%iles using the 2~$mm$ cube average. The electric field within the non-skin tissues also remains consistent within 99.9 to 99.0~\%iles. A more significant electric field value increase is observed using the 5~$mm$ line average in both skin and non-skin tissues. In XCAT model, the electric field values are almost the same at 99~\%ile in both hand postures considering both averaging techniques. In higher percentiles, 2-5 times values of electric field in skin are observed in close loop position. Alternative to what observed above in the non-skin tissues, we found the increase of 3-16 times in non-skin tissues considering the two different hand postures (open/close loops).

%-----------------------
% Table 2
%-----------------------

\begin{table*}
\centering
\footnotesize
\caption{Percentiles of 2~$mm$ cube- and 5~$mm$ line-averaged electric field for the first scenario ($\times 10^{-2}$ V/m).}
\label{Tab2}
\setlength{\tabcolsep}{3pt}
\begin{tabular}{|c|c|c|c|c|c|c|c||c|c|c|c|c|c|c|}
\hline
\multirow{3}{*}{Percentile [\%]} &\parbox[t]{2mm}{\multirow{3}{*}{\rotatebox[origin=c]{90}{Avg.}}}& \multicolumn{6}{c||}{Open hand mode} & \multicolumn{6}{c|}{Close hand mode}\\
\cline{3-14}
 && \multicolumn{3}{c|}{TARO} & \multicolumn{3}{c||}{XCAT}& \multicolumn{3}{c|}{TARO} & \multicolumn{3}{c|}{XCAT}\\
\cline{3-14}
 && All tissues & Skin & Others & All tissues & Skin & Others & All tissues & Skin & Others & All tissues & Skin & Others\\
\hline \hline
100.00 &\parbox[t]{2mm}{\multirow{3}{*}{\rotatebox[origin=c]{90}{2~mm$^3$}}} &0.346 & 0.346 & 0.285 & 0.456 & 0.456 & 0.234 & 3.873 & 3.873 & 1.714 &2.121 & 2.121 & 1.011 \\
\cline{1-1}\cline{3-14}
~99.90 & &0.214 & 0.278 & 0.131 & 0.164 & 0.364 & 0.152 & 1.509 & 2.375 & 0.264 & 1.564 & 1.970 & 0.307 \\
\cline{1-1}\cline{3-14}
~99.00 & &0.119 & 0.199 & 0.099 & 0.114 & 0.202 & 0.112 & 0.334 & 1.393 & 0.106 & 0.314 & 1.743 & 0.124 \\
\hline \hline
100.00 &\parbox[t]{2mm}{\multirow{3}{*}{\rotatebox[origin=c]{90}{5~mm}}}& 0.297 & 0.297 & 0.220 & 0.309 & 0.310 & 0.200 & 2.174 & 2.174 & 1.929 & 1.039 & 1.039 &0.903 \\
\cline{1-1}\cline{3-14}
~99.90 && 0.194 & 0.236 & 0.125 & 0.181 & 0.237 & 0.154 & 1.387 & 1.642 & 0.991 & 0.898 & 0.984 & 0.700 \\ 
\cline{1-1}\cline{3-14}
~99.00 && 0.123 & 0.155 & 0.094 & 0.123 & 0.165 & 0.111 & 0.614 & 1.046 & 0.256 & 0.559 & 0.856 & 0.266 \\
\hline  
\end{tabular}
\end{table*}

%-----------------------
% Table 3
%-----------------------

\begin{table*}
\centering
\footnotesize
\caption{Percentiles of 2~$mm$ cube- and 5~$mm$ line-averaged electric field for the second scenario ($\times 10^{-2}$ V/m).}
\label{Tab3}
\setlength{\tabcolsep}{3pt}
\begin{tabular}{|c|c|c|c|c|c|c|c||c|c|c|c|c|c|}
\hline
\multirow{3}{*}{Percentile [\%]} &\parbox[t]{2mm}{\multirow{3}{*}{\rotatebox[origin=c]{90}{Avg.}}}& \multicolumn{6}{c||}{Open loop mode (NonTouching)} & \multicolumn{6}{c|}{Close loop mode (Touching)}\\
\cline{3-14}
 && \multicolumn{3}{c|}{TARO} & \multicolumn{3}{c||}{XCAT}& \multicolumn{3}{c|}{TARO} & \multicolumn{3}{c|}{XCAT}\\
\cline{3-14}
 && All tissues & Skin & Others & All tissues & Skin & Others & All tissues & Skin & Others & All tissues & Skin & Others\\
\hline \hline
100.00 &\parbox[t]{2mm}{\multirow{3}{*}{\rotatebox[origin=c]{90}{2~mm$^3$}}}& 0.378 & 0.225 & 0.378 & 0.604 & 0.604 & 0.372 & 6.337 & 6.337 & 5.862 & 5.295 & 5.295 & 1.942 \\
\cline{1-1}\cline{3-14}
~99.90 && 0.130 & 0.205 & 0.130 & 0.227 & 0.405 & 0.212 & 0.103 & 0.154 & 0.102 & 0.237 & 2.822 & 0.217\\
\cline{1-1}\cline{3-14}
~99.00 && 0.070 & 0.137 & 0.070 & 0.099 &0.254 & 0.094 & 0.061 & 0.112 & 0.061 & 0.101 & 0.303 & 0.095\\
\hline 
\hline
100.00 &\parbox[t]{2mm}{\multirow{3}{*}{\rotatebox[origin=c]{90}{5~mm}}}& 0.508 & 0.508 & 0.466 & 0.497 & 0.497 & 0.361 & 7.299 & 7.299 & 6.932 & 2.495 & 2.490 & 2.495\\
\cline{1-1}\cline{3-14}
~99.90 &&  0.143 & 0.207 & 0.130 & 0.266 & 0.373 & 0.211 & 0.183 & 0.561 & 0.163 & 0.338 & 1.402 & 0.237\\
\cline{1-1}\cline{3-14}
~99.00 && 0.076 & 0.126 & 0.070 & 0.116 & 0.220 & 0.093 & 0.092 & 0.147 & 0.086 & 0.124 & 0.266 & 0.096\\
\hline 
\end{tabular}
\end{table*}

%-----------------------
% V. Discussion
%-----------------------

\section{Discussion}

In this study, first, an experimental study was designed for a potential change of the electrostimulation threshold by the skin-to-skin contact. This is based on some recent computational studies which suggest the \emph{in situ} electric field becomes high due to the skin-to-skin contact in the anatomical models. Instead, the IEEE C95.1-2019 suggested that the electric field at the contact part should be excluded for compliance because it is not related to the stimulation. However, evidence may not be enough, especially for specific scenarios where equivalent exposed area becomes large due to skin-to-skin contact. 

The scenarios in this study were designed so that the enhancement of the electric field due to the skin-to-skin is expected in the computation. If the measured electrostimulation threshold was changed by the skin-to-skin contact, the description on the standard should be revisited. If not, the high electric field at the skin-to-skin contact is attributable to poor modeling of the skin layer, which is not related to an adverse health effect. 

In the experiment, the main emphasis was that the stimulation hand/finger part by the external magnetic field was carefully considered, unlike previous studies. This was based on the fact of a 4:1 variation in the electrical threshold of perception at different locations of the body \cite{Reilly1998} (Fig. 7.18, pp. 276). In previous magnetic resonance (MR) studies \cite{DenBoer2002JMRI}, comparison of the two conditions is thus complicated by the fact that the threshold sensitivity to the electrostimulation varies substantially over different parts of the body. A potential 30\% difference in the threshold reported in previous studies might reflect that the stimulation site would be different for different exposure conditions \cite{Reilly2018private}.
  
Our experimental study suggested that the electrostimulation threshold was almost identical for two scenarios. Instead, a higher computed electric field was observed in the part of the skin-to-skin contact. This experimental and computational comparison suggests that the weakness of the skin modeling is the primary reason for this variation. The skin is comprised of multi-layers, hypodermis, dermis, and epidermis. Note that the epidermis is the outermost of the three layers and is further classified into additional sub-layers (stratum corneum, stratum lucidum,\dots, etc). In addition, the skin has anisotropic conductivity; specifically, the conductivity in the transverse direction is high whereas low in the depth direction \cite{Yamamoto1977MBEC, Yamamoto1978MBEC}. Thus, it may not be sufficient to model the skin as a single tissue with a millimeter resolution or single conductivity. Instead, it would be unpractical to model the human body with a higher resolution because tissue conductivity has not been measured accurately for different tissues (e.g., the thickness of the stratum corneum is the order of 20 $\mu$m) \cite{Gabriel2009PMB, Wake2016PMB}.
 
The limitation of this experimental study was that the skin condition was dry. Thus, the exposure for the wet skin (e.g., heavily sweating) is not considered. In such cases, the conductivity itself may change dramatically. However, this experimental condition may not be feasible because the skin conductivity does not readily change its value even if the hand is immersed in the saline solution. For example, it takes a few minutes for the skin impedance to reach a constant value after the application of the voltage in transcranial direct stimulation~\cite{GomezTames2016PMB}. Considering subject variability, the skin thickness at the skin-to-skin touch region was not measured. However, we have reported that induced electrical field for TMS marginally influenced by the skin thickness in human head models \cite{Rashed2019Access}.

According to IEEE C95.1 standard~\cite{IEEEC9512019}, it has been mentioned that ``\emph{The safety limits for electrostimulation are based on conservative assumptions of exposure. However, they cannot address every conceivable assumption}". Thus, the protocol and/or the scenario considered here is sufficient on the skin-to-skin contact. Rather, the systematic measurement and computation supported its description that ``\emph{Dosimetric reference limits do not apply to induced fields or current crossing skin-to-skin contact}."

%-----------------------
% VI. Conclusion
%-----------------------

\section{Conclusion}

The international standards/guidelines are designed for human from potential adverse health effect. This study, despite the limited cases but considering the generality based on the Faraday's law, suggested that the skin-to-skin contact does not result in lowering the threshold experimentally. In this study, we introduce a kinematic method for hand modeling for different postures setup. The proposed method is used to adjust hand models extracted from TARO and XCAT models to simulate different scenarios of skin-to-skin contacts. TARO model is extensively in electromagnetic computation studies, however, XCAT was hardly being used, though it provides an interesting anatomy representation. The computational results were not consistent, where the resolution of skin is inherently insufficient. However, to consider the skin model with a resolution of micrometer would be impractical. Our finding supports that the skin-to-skin contact should be excluded as mentioned in the IEEE C95.1-2019. Our finding may also suggest that the skin-to-skin contact is not essential for protecting human, rather a computational modeling issue, which may potentially be related to the evaluation of compliance assessment in the product safety, where how to deal with the skin-to-skin contact. More detailed description on how to deal with skin-to-skin contact would be needed for reproducible compliance assessment.

\bibliographystyle{IEEEbib}
\bibliography{IEEEabrv,Refs1}
\end{document}